# Status Report for the RHIC Control System*

J. Morris, T. D'Ottavio
BNL, Upton, NY 11973

Abstract

The Relativistic Heavy Ion Collider (RHIC) at Brookhaven has completed nearly two years of successful commissioning and operation. The control system is briefly reviewed and its contribution to the RHIC effort is analyzed, with emphasis on special challenges and innovative design: continuing efforts are also discussed.

## 1 INTRODUCTION

The Relativistic Heavy Ion Collider (RHIC) consists of two concentric rings with a circumference of 2.4 miles. The rings are made up of 1,740 superconducting magnets. The RHIC control system must address the special challenges posed by the size and character of RHIC. Precise beam control is needed to avoid quenches of cryogenic magnets and to provide adequate beam lifetime for stores of 4 hours or more. Control of settings for the widely distributed magnets must be well synchronized. Data acquisition from beam instrumentation must also be synchronized. Data capture and diagnostic tools are needed to support analysis of acceleration ramps and post mortem analysis of beam aborts and magnet quenches. During the past two years, the RHIC control system has continued to evolve to meet the demands of accelerator commissioning and operations.

## 2 SYSTEM OVERVIEW

### 2.1 Controls Hardware

The RHIC control system consists of two physical levels: console level computers and front-end computers (FECs). FECs provide access to accelerator equipment. FECs consist of a VME chassis with a single board computer, network connection, and I/O modules. Most FECs use the Motorola PowerPC™ processor. Commercially available VME I/O modules are used when possible. VME modules have been custom-designed for functions including power supply control, signal acquisition, and timing.

Console level computers are Sun™ workstations. One prototype Linux console has been used in the main control room during the 2001 RHIC run. Xterminals are also used as consoles in RHIC service buildings and other remote locations. The console level includes server machines that host processes for services such as data logging and alarming. Dedicated servers provide database and file services.

Console level computers and FECs are networked via Ethernet. A number of data links[1] provide synchronization of operations. The links are the Event Link, the Real Time Data Link[2], the Permit and Quench Interlock Links, and two Beam Synchronous Links.

### 2.2 Controls Software

Front-end computers (FECs) use the VxWorks™ real-time operating system. FEC software modules fall into two broad categories: device drivers and accelerator device objects (ADOs[3]). Device drivers, typically written in C, simply provide a software interface to the VME I/O modules. ADOs are C++ objects that are responsible for coordinating operation of the accelerator equipment and providing an interface for client programs. A code generation mechanism facilitates the development of ADO code and associated database information.

Console level software runs under the Solaris™ operating system. Console software has been primarily developed in the C++ language. Programmers rely on an extensive set of C++ class libraries. Members of both Controls and Physics groups develop RHIC applications. Many key physics applications have been written in Tool Command Language (TCL)[4].

The UI Toolkit is a large C++ class library that provides an object-oriented interface to X/Motif widgets and third party graphics and table tools. A C++ class interface is provided for access to Sybase database services. Assorted data storage classes are used to read/write Self Describing Data Set (SDDS)[5] files, often with associated database information to simplify searches for files of interest. C++ classes have been written to provide an application interface to accelerator device objects. A Common DEVice (CDEV)[6] interface is also offered. The CDEV interface is available to C++ programs but it also provides control system access for TCL programs.

Server processes can be built from the same ADO tools used on FECs and present the same ADO interface to application programs. Other servers are built in the CDEV generic server style.

## 3 SYSTEM HIGHLIGHTS

This section describes some of the aspects of RHIC controls that have been most significant for the current run.

### 3.1 Model Driven Ramp Control

Model driven control of magnet ramps[7] is a well established part of RHIC operations. The path up the ramp follows *stepstones*, which are points at which the optics of the accelerator is specified. Physical properties of the machine, like tune and chromaticity, can be specified at these *stepstones*. An on-line model server converts these properties to magnet strengths.

{*} Work supported by US Department of Energy

## 3.2 Sequencing

Over the last year, a significant effort was put into developing a software infrastructure that supports the sequencing of routine operational procedures[8]. This system is now used for a wide variety of purposes both within RHIC systems and the injector accelerator systems that feed RHIC. General areas of use include setting up and checking instrumentation and power supply systems, ramping RHIC through energy states, and system recovery from fault states.

The sequences themselves are written in a simple sequencer language developed at BNL. Each step in a sequence is either a primitive task (set a value, trigger an event, etc.) or a call to another sequence. The nesting capability enables reusability of sequences and permits more complex sequences to be constructed. No logic or parallel execution is currently allowed in the sequences, although these enhancements are being considered for a future version.

The sequencing software infrastructure includes two different GUI applications and two server programs. The two GUI programs service different sets of users, while the server programs allow sequences to be run by other applications and servers. The execution of all sequences is logged, allowing for easier diagnostics when problems arise.

The sequencing system has been heavily used over the last year and has lived up to its initial expectations: reproducible playback of procedures, improved execution times, and minimization of errors. To date, over 100 sequences have been created to handle routine operations of the RHIC complex.

## 3.3 Power Supply Diagnostics

Successful operation of RHIC depends on reliable and reproducible performance of more than 800 power supplies for ring magnets during acceleration ramps. A system was developed to capture detailed power supply measurements from RHIC ring power supplies during ramps. Diagnostic tools were developed to support analysis of this ramp data [9].

The ramp data sets provide a quantitative measure of the reproducibility of power supply performance. They are used to analyze power supply response when new modes of ramping are introduced and to identify power supply problems that may go undetected by ordinary alarm mechanisms. The ramp diagnostic system has now become a routine part of RHIC operations. Data is captured during every ramp and routinely analyzed by control room personnel. A watching mechanism has also been introduced to bring problems to the immediate attention of operators.

## 3.4 Coordinated Data Acquisition & Storage

In the fall of 2000, a project was initiated to improve the Control System data logging and retrieval capabilities. The goal of this project was to ensure that data needed for the analysis of RHIC could be captured and stored in a way that allowed easy retrieval and correlation of data from all systems. This project built on earlier work to establish data correlation[10] and logging infrastructure. New effort was focused in the following areas.

1) *Coordinated Acquisition* - acquire data from all necessary systems at appropriate rates and ensure that acquisition of data from different systems is synchronized to the level required for correlation. Standard events were defined to trigger data acquisition and systems were modified as necessary to conform to standard triggering methods. Since appropriate acquisition rates for some systems change during the course of a machine cycle (e.g. much more data is typically needed during ramps), the Sequencer was used to put these data acquisition triggers in the proper mode for each machine state.

2) *Data Storage* – ensure that adequate data and time stamp information is stored for all systems. Upgrades to the logging system were necessary to support some types of data and to maintain sufficient timing information for correlation. The common Logger format (based on SDDS) is used for almost all systems. Additional changes have been made to simplify access to data. Database headers, stored for each data file, are tagged with a *fill number*. A *fill* is defined as the period of time encompassing injection, acceleration, and storage of beam in RHIC. Files are also stored in a directory structure by *fill*, a procedure that has long been in use at CERN's LEP facility[11].

3) *Data Retrieval* - provide simple mechanisms for display of data from multiple systems with data selected by time period or *fill number*. Graphic display programs have been upgraded to add selection of data by *fill number* and to accommodate the display of array data in assorted formats. Additional enhancements include the ability to add event markers to graphs. The ability to plot data relative to an event allows the overlay of data from different fills.

The improvements have been extremely important to RHIC running in 2001. Data is being routinely logged for all systems. The viewing tools are routinely used to analyze machine performance.

## 3.5 Post Mortem

The RHIC Post Mortem (PM) system[12] was designed and developed to provide information about the state of the collider at the time of a beam abort, a quench of one of the RHIC superconducting magnets, or some other failure event that may cause beam to be lost from the machine. The data collected by this system helps to determine 1) the cause of the failure, 2) whether the machine is ready for another injection, and 3) how future stores might be improved.

Events on the RHIC event link (e.g. beam abort, quench) are the triggers that cause data, buffered for this purpose, to be read and stored by software systems that exist on both FECs and console-level computers. A GUI application then allows a user to view and filter the available data by system (e.g. power supply, beam loss monitor) and to select the data of interest, which is displayed graphically in a system-specific way.

PM data is currently gathered from the power supply, beam loss monitor, current transformer, quench detection and real-time data link systems. The PM system was used extensively as a diagnostic tool by the power supply/magnet groups throughout RHIC's first two years and continues to serve that purpose.

*3.6 Tune Feedback*

A tune feedback system [13] has been undergoing commissioning. Variations in measured tune are captured by a phase locked loop tune measurement system. Compensating changes in magnet strength are calculated in the tune measurement FEC and delivered to the Real Time Data Link (RTDL) via a reflective memory connection. The RTDL values are used to make appropriate adjustments in magnet power supply references.

## 4 PERFORMANCE/RELIABILITY

A significant and steady effort has been devoted to maintaining reliable operation of the control system during RHIC commissioning. Effort has also been directed at ensuring that failures are reported promptly by the alarm system and that recovery can be accomplished with minimal disruption to accelerator operation. During the first physics run in the summer of 2000, control system failures contributed less than 2% of RHIC downtime. The statistics are expected to be similar for 2001. A significant source of controls failures in 2001 has been radiation induced memory upsets in equipment alcoves located near the RHIC tunnel. Logging and post mortem servers have had a high degree of reliability but occasional failures have resulted in lost data. Graphic display performance is sometimes a bottleneck, particularly on Xterminal displays.

## 5 FUTURE

Final commissioning of the tune feedback system is anticipated before the end of the current RHIC run. To minimize recovery time after quench protection interlocks, software is being developed for automated analysis of post mortem data. Alcove radiation problems are being analyzed and solutions considered, including the relocation of some sensitive equipment.

Work to increase the reliability of logging and post mortem servers will continue. Work will be undertaken to improve graphic display performance. Disk storage space will be expanded to accommodate the demands of the post mortem, power supply diagnostic and beam instrumentation systems. The use of Linux consoles will be expanded in the Main Control Room. The Linux consoles are more economical than Sun workstations and may be outfitted with four monitors. A pilot project is underway to evaluate the use of the Java language for future software development.

TUAT002

## ACKNOWLEDGEMENTS

The authors gratefully acknowledge the contributions of Don Barton, Jonathan Laster, Robert Michnoff, and Johannes van Zeijts.

that exist on both FECs and console-level computers. A GUI application then allows a user to view and filter the available data by system (e.g. power supply, beam loss monitor) and to select the data of interest, which is displayed graphically in a system-specific way.

PM data is currently gathered from the power supply, beam loss monitor, current transformer, quench detection and real-time data link systems. The PM system was used extensively as a diagnostic tool by the power supply/magnet groups throughout RHIC's first two years and continues to serve that purpose.

*3.6 Tune Feedback*

A tune feedback system [13] has been undergoing commissioning. Variations in measured tune are captured by a phase locked loop tune measurement system. Compensating changes in magnet strength are calculated in the tune measurement FEC and delivered to the Real Time Data Link (RTDL) via a reflective memory connection. The RTDL values are used to make appropriate adjustments in magnet power supply references.

## 4 PERFORMANCE/RELIABILITY

A significant and steady effort has been devoted to maintaining reliable operation of the control system during RHIC commissioning. Effort has also been directed at ensuring that failures are reported promptly by the alarm system and that recovery can be accomplished with minimal disruption to accelerator operation. During the first physics run in the summer of 2000, control system failures contributed less than 2% of RHIC downtime. The statistics are expected to be similar for 2001. A significant source of controls failures in 2001 has been radiation induced memory upsets in equipment alcoves located near the RHIC tunnel. Logging and post mortem servers have had a high degree of reliability but occasional failures have resulted in lost data. Graphic display performance is sometimes a bottleneck, particularly on Xterminal displays.

## 5  FUTURE

Final commissioning of the tune feedback system is anticipated before the end of the current RHIC run. To minimize recovery time after quench protection interlocks, software is being developed for automated analysis of post mortem data. Alcove radiation problems are being analyzed and solutions considered, including the relocation of some sensitive equipment.

Work to increase the reliability of logging and post mortem servers will continue. Work will be undertaken to improve graphic display performance. Disk storage space will be expanded to accommodate the demands of the post mortem, power supply diagnostic and beam instrumentation systems. The use of Linux consoles will be expanded in the Main Control Room. The Linux consoles are more economical than Sun workstations and may be outfitted with four monitors. A pilot project is underway to evaluate the use of the Java language for future software development.

## ACKNOWLEDGEMENTS

The authors gratefully acknowledge the contributions of Don Barton, Jonathan Laster, Robert Michnoff, and Johannes van Zeijts.

## REFERENCES

[1] B.R. Oerter, "Accelerator Timing at the Relativistic Heavy Ion Collider," Proc. ICALEPCS'99, Trieste, p.191
[2] H. Hartmann, "The RHIC Real Time Data Link System," Proc. 1997 IEEE Particle Accelerator Conf., Vancouver, p.2499
[3] L.T. Hoff, J.F. Skelly, "Accelerator Devices as Persistent Software Objects," Proc. ICALEPCS'93, Berlin, Nucl. Instr. and Meth. A 352(1994)185
[4] Ousterhout, J.K. ,"Tcl and the Tk Toolkit", Addison-Wesley Professional Computing, 1994
[5] M. Borland, "A Self Describing File Protocol for Simulation Integration and Shared Postprocessors," Proc. 1995 IEEE Particle Accelerator Conf., Dallas, p.2184
[6] J. Chen et al., "CDEV: An Object-Oriented Class Library for Developing Device Control Applications," Proc. ICALEPCS'95, Chicago, p.97
[7] J. van Zeijts, "Model Driven Ramp Control at RHIC", this conference
[8] T. D'Ottavio "Description of the RHIC Sequencer", this conference
[9] J. T. Morris, "Power Supply Ramp Diagnostics", this conference
[10] R. Michnoff et al., "RHIC Data Correlation Methodology," Proc. 1999 IEEE Particle Accelerator Conf., New York, p.693
[11] M. Lamont, CERN, private correspondence
[12] J.S.Laster, "Post Mortem System – Playback of the RHIC Collider", this conference
[13] J. van Zeitjs, "Tune Feedback at RHIC", this conference